\documentstyle[mprocl]{article}

\def\Journal#1#2#3#4{{#1} {\bf #2}, #3 (#4)}

\def\JMP{{\em J. Math. Phys.}}
\def\CQG{{\em Class. Quantum Gravit.}}

\begin{document}

\title{HAMILTONIANS FOR COMPACT HOMOGENEOUS UNIVERSES}

\author{ Masayuki TANIMOTO }

\address{Yukawa Institute for Theoretical Physics, 
        Kyoto University, Kyoto, Japan}

\author{ Tatsuhiko KOIKE }

\address{Department of Physics, Keio University, Kanagawa, Japan}

\author{ Akio HOSOYA }

\address{Department of Physics, Tokyo Institute of Technology,
Tokyo, Japan}

\maketitle

\abstracts{ We briefly show how we can obtain Hamiltonians for spatially
  compact locally homogeneous vacuum spacetimes. The dynamical variables
  are categorized into the curvature parameters and the Teichm\"{u}ller
  parameters. While the Teichm\"{u}ller parameters usually parameterise the
  covering group of the spatial sections, we utilise another suitable
  parameterization where the universal cover metric carries all the
  dynamical variables and with this we reduce the Hamiltonians.  For
  our models, all dynamical variables possess their clear geometrical
  meaning, in contrast to the conventional open models.}


\def\d{{\rm d}}
\def\m{{\bf m}}
\def\n{{\bf n}}
\def\r{{\bf r}}
\def\v{{\bf v}}
\def\x{{\bf x}}
\def\R{{\bf R}}
\def\Z{{\bf Z}}
\def\del{\partial}
\def\Lap{\bigtriangleup}
\def\^{\wedge}
\def\goinf{\rightarrow\infty}
\def\goes{\rightarrow}
\def\bm{\boldmath}
\def\-{{-1}}
\def\inv{^{-1}}
\def\sqr{^{1/2}}
\def\isqr{^{-1/2}}

\def\reff#1{(\ref{#1})}
\def\vb#1{{\partial \over \partial #1}} 
\def\Del#1#2{{\partial #1 \over \partial #2}}
\def\Dell#1#2{{\partial^2 #1 \over \partial {#2}^2}}
\def\Dif#1#2{{d #1 \over d #2}}
\def\Lie#1{ {\cal L}_{#1} }
\def\diag#1{{\rm diag}(#1)}
\def\abs#1{\left | #1 \right |}
\def\rcp#1{{1\over #1}}
\def\paren#1{\left( #1 \right)}
\def\brace#1{\left\{ #1 \right\}}
\def\bra#1{\left[ #1 \right]}
\def\angl#1{\left\langle #1 \right\rangle}
\def\lvector#1#2#3#4{\paren{\begin{array}{c} #1 \\ #2 \\ #3 \\ #4 \end{array}}}
\def\vector#1#2#3{\paren{\begin{array}{c} #1 \\ #2 \\ #3 \end{array}}}
\def\svector#1#2{\paren{\begin{array}{c} #1 \\ #2 \end{array}}}
\def\matrix#1#2#3#4#5#6#7#8#9{
        \left( \begin{array}{ccc}
                        #1 & #2 & #3 \\ #4 & #5 & #6 \\ #7 & #8 & #9
        \end{array}     \right) }
\def\smatrix#1#2#3#4{
        \left( \begin{array}{cc} #1 & #2 \\ #3 & #4 \end{array} \right) }
\def\manbo#1{}  

\def\diffeos{diffeomorphisms}
\def\diffeo{diffeomorphism}
\def\Teich{{Teichm\"{u}ller }}

\def\bm{\boldmath}
\def\Gam{\mbox{$\Gamma$}}
\def\d{{\rm d}}

\def\hh{{h}}
\def\gg{{\rm g}}
\def\uh#1#2{\hh^{#1#2}}
\def\dh#1#2{\hh_{#1#2}}
\def\ug#1#2{\gg^{#1#2}}
\def\dg#1#2{\gg_{#1#2}}
\def\uug#1#2{\tilde{\gg}^{#1#2}}
\def\udg#1#2{\tilde{\gg}_{#1#2}}
\def\udh#1#2{\tilde{\hh}_{#1#2}}
\def\ustd#1#2{\tilde{h}_{#1#2}^{\rm std}}

\def\MM{\four{M}}
\def\uMM{\four{\tilde{M}}}
\def\Mtil{\tilde M}

\def\Mt{\mbox{$M_t$}}
\def\uMt{\mbox{$\tilde{M}_t$}}
\def\uMi{\mbox{$\tilde{M}_{t_0}$}}
\def\UC{\mbox{$(\uMM,\udg ab)$}}
\def\CU{\mbox{$(\MM,\dg ab)$}}
\def\UCh{\mbox{$(\uMt,\udh ab)$}}
\def\UCs{\mbox{$(\Mtil,\udh ab)$}}
\def\UChi{\mbox{$(\uMti,\udh ab)$}}
\def\CUh{\mbox{$(\Mt,\dh ab)$}}

\def\tei{\mbox{\bm $\tau$}}
\def\cur{\mbox{\bm $r$}}
\def\tcv{\mbox{$(\cur, \tei, v)$}}
\def\u{{\bf u}}
\def\g{{\bf g}}
\def\h{{\bf h}}
\def\std{{\rm std}}
\def\dyn{{\rm dyn}}
\def\UCx#1{(\uMM,\udg ab{#1})}
\def\CUx#1{(\MM,\dg ab{#1})}
\def\gdy#1{\udg ab^\dyn{#1}}
\def\UCgdyx#1{(\uMM,\gdy{#1})}
\def\UCu{\mbox{$\UCx{[\u]}$}}
\def\CUug{\mbox{$\CUx{[\u,\g]}$}}
\def\UCgdycurtei{\mbox{$\UCgdyx{(\cur,\tei)}$}}
\def\hst#1{\udh ab^\std{#1}}
\def\hdy#1{\udh ab^\dyn{#1}}
\def\UChx#1{(\uMt,\udh ab{#1})}
\def\UCstx#1{(\Mtil,\hst{#1})}
\def\UCdyx#1{(\Mtil,\hdy{#1})}
\def\UChdyx#1{(\uMt,\hdy{#1})}
\def\UChu{\mbox{$\UChx{[\u]}$}}
\def\UCst{\mbox{$\UCstx{}$}}
\def\UCstcur{\mbox{$\UCstx{[\cur]}$}}
\def\UCdycurtei{\mbox{$\UCdyx{[\cur,\tei]}$}}
\def\UChdycurtei{\mbox{$\UChdyx{[\cur,\tei]}$}}

\def\s#1{\sigma^{#1}}
\def\a#1#2{a_{#1}{}^{#2}}
\def\p#1#2{p^{#1}{}_{#2}}
\def\dug#1#2{g_{#1}{}^{#2}}
\def\H#1#2{H_{#1#2}}
\def\f#1#2{f^{#1}{}_{#2}}

\section{Introduction}

Spatially homogeneous cosmological models ${}^{\rm e.g.,}$ \cite{RS} give
good prototypes for many theoretical models in gravity and cosmology.
Some controversies arise, however, when considering Hamiltonian
structures of them, which are of large interest, especially, for an
application to quantum gravity.  For example, it is well known \cite{RS}
that the models known as Bianchi class B do not possess a natural
Hamiltonian reduced from the full Hamiltonian.  Even for the class A
models, a sort of discrepancies of dynamical degrees of freedom is
pointed out by Ashtekar and Samuel \cite{AS}. For example, the Kasner
solution, which is the vacuum solution of Bianchi I, has only one
dynamical degree of freedom, i.e., there is only one free parameter
which can be specified freely at an initial Cauchy surface.  Note,
however, that an odd number of dynamical degrees of freedom cannot come
out from a Hamiltonian system.  So, when one wants a Hamiltonian, one
usually works with the so-called diagonal model, which has three
dynamical variables and gives four dynamical degrees of freedom in the
Hamiltonian view.  If we work with the full, nondiagonal model, which
may be the most natural in the Hamiltonian view, we have ten dynamical
degrees of freedom with six dynamical variables. Thus, we have obtained
three possible numbers (i.e., 1, 4 and 10) of dynamical degrees of
freedom for Bianchi I!

These discrepancies for the spatially {\it open} model are responsible
for the ambiguous specification of the dynamical variables.  Note that a
(group invariant) spatial section of the spatially open Bianchi I model
is a three dimensional Euclid space, which has no free parameters
specifying the intrinsic geometry, since any Euclid space is isometric
to the one with the standard metric $\d x^2+\d y^2+\d z^2$.  This proves
that the open Bianchi I model possesses no dynamical variables(, though
it has one ``dynamical degrees of freedom'' as in the Kasner solution).
When employing the diagonal or full model, the metric components are
treated as if they are true dynamical variables, but in this case they
lose their geometrical nature.  As a result, the open Bianchi I model
cannot admit a consistent Hamiltonian structure.

How about spatially {\it compact} Bianchi models?  For example, we can
compactify the Euclid space and make a torus $T^3$ by first fixing three
independent vectors $\vec{a_1}$, $\vec{a_2}$, and $\vec{a_3}$ in the
standard metric and then identifying each two points $\vec p$ and $\vec
q$ such that $\vec p-\vec q=k \vec{a_1}+l \vec{a_2}+m \vec{a_3}$,
$k,l,m\in\Z$.  If we smoothly vary the three vectors $\vec{a_i}$, then
the quotient manifold in general smoothly varies nonisometrically.  More
precisely, six independent parameters in $\vec{a_i}$ can induce
nonisometric deformations of the quotient.  Such parameters, denoted
collectively as \tei, are called {\it \Teich parameters} \cite{Th,Oh}.
We may regard the \Teich parameters as (part of the) dynamical variables
of a spatially compactified locally homogeneous spacetime.  As for the
$T^3$ model on Bianchi I, the dynamical variables are the six \Teich
parameters only, and we can prove for this system there exists a
consistent Hamiltonian structure.

In this article, we show a skeleton of our method \cite{KTH,TKH1,TKH2}
of obtaining consistent Hamiltonians for spatially compact locally
homogeneous (SCH) spacetimes.  (See also the paper by Kodama \cite{Kod},
where a somewhat different approach is presented.)

\section{Method for reducing the Hamiltonian
 for a compact homogeneous universe}

As mentioned above, a flat torus is locally isometric to the standard
Euclid space $(\R^3,\eta_{ab})$, where $\eta_{ab}$ is the
standard Euclid metric.  The six \Teich parameters $\tei=\{\a ij\}$ are
the independent parameters in the covering group $A_{\tei}$, represented
by
\begin{equation}
  A_{\tei}=\brace{\vec a_1,\vec a_2,\vec a_3}=
  \brace{\vector{\a11}00, \vector{\a21}{\a22}0, \vector{\a31}{\a32}{\a33}}.
        \label{1}
\end{equation}
The flat torus is then represented by $(\R^3,\eta_{ab})/A_{\tei}.$
{\it Similarly}, any compact and locally homogeneous manifold can be
represented by
\begin{equation}
        (\tilde M,\ustd ab[\cur])/A_{\tei},
        \label{3}
\end{equation}
where $(\tilde M,\ustd ab[\cur])$ is a homogeneous manifold which is
free from \diffeos, $\cur$ is a set of parameters in a standard metric
$\ustd ab$, and $\tei$ is a set of \Teich parameters.  A proper
action of $A_{\tei}$ on $\tilde M$ is understood.  The parameters
$(\cur,\tei)$ {\it are} our dynamical variables.  (For the flat torus case,
$\cur=\emptyset$.)

We then define a \diffeo\
$\phi_{\tei}:\;\Mtil\goes\Mtil$ such that
\begin{equation}
        A_{\tei}=\phi_{\tei}\circ A_0\circ\phi_{\tei}\inv,
        \label{3-1}
\end{equation}
where $A_0$ is the covering group for a set of fixed \Teich parameters
$\tei=\tei_0$. 
With this, we obtain another parameterization
\begin{equation}
        \UCdycurtei/A_0,
        \label{par2}
\end{equation}
where
\begin{equation}
        \hdy{[\cur,\tei]}\equiv\phi_{\tei*}\hst{[\cur]}.
        \label{3-2}
\end{equation}
We shall refer to $\phi_{\tei}$ as a {\it \Teich \diffeo} (TD).  TDs are
not unique. We refer to a TD implemented in the HPDs \cite{AS} as an
{\it HPTD}.

If we consider a spacetime metric $\gdy{(\cur,\tei)}$ whose spatial part
is given by Eq.\reff{3-2}, where $\cur$ and $\tei$ are free functions of
time $t$, then this gives a possible ansatz of a SCH spacetime.  We do
this with the HPTDs and the synchronous gauge, since the ansatz obtained
by doing so, an ordinary Bianchi type spacetime metric, gives a
dynamically consistent one, since the extrinsic curvature and the metric
contains the same transitive symmetry (isometry) group.

As for $T^3$ on Bianchi I, the HPTD is given by the following
linear transformation
\begin{equation}
        \phi_{\tei}: \vector xyz \goes
        \matrix{\a11}{\a21}{\a31}0{\a22}{\a32}00{\a33} \vector xyz .
        \label{a2-Mtei}
\end{equation}
We can calculate the spatial metric according to Eq.\reff{3-2} by inducing
the standard Euclid metric, and finally obtain the spacetime metric with 
the synchronous gauge with $\a ij$ being functions of time.
The Hamiltonian can be reduced with this spacetime metric ansatz.

\section{Concluding remark}

We have very shortly shown how we obtain Hamiltonians for spatially
compact locally homogeneous spacetimes.  The points are the use of the
parameterization \reff{par2} and the use of the HPTDs.
The spacetime metric ansatz thereby obtained enables us to reduce the
Hamiltonian for many spatially compact locally homogeneous spacetime
models.  We remark, however, that in some cases the HPTDs exist only for
part of the \Teich deformations, or do not exist at all.  In these
cases, the dynamics of the \Teich deformations degenerates or freezes,
respectively.\cite{TKH2}

\section*{Acknowledgments}
We acknowledge financial support from the Japan Society for the Promotion
of Science (M.T.) and the Ministry of Education, Science and Culture
(M.T. \& T.K.).

\end{document}